\documentclass[preprint,prd,aps,article,nofootinbib]{revtex4}
\usepackage{hyperref}
\usepackage{color}
\usepackage{slashed}
\usepackage{amsmath}
\allowdisplaybreaks[1]
\usepackage{graphicx}
\usepackage{comment}
\usepackage{epstopdf}
\usepackage{subfigure}
\usepackage{verbatim}

\begin{document}
\title{Dalitz decays of vector heavy quarkonia into $\chi_{QJ}(1P)$ in the Bethe–Salpeter approach}
\author{Xin-Wen Wang$^{1,2,3}$,
	Wen-Yuan Ke$^{1,2,3}$,
	Su-Yan Pei$^{4}$,
    Tianhong Wang$^5$,
    Qiang Li$^6$,
	and Guo-Li Wang$^{1,2,3}$\footnote{Corresponding author}}
\affiliation{$^1$ Department of Physics, Hebei University, Baoding 071002, China\\
	$^2$ Hebei Key Laboratory of High-precision Computation and
	Application of Quantum Field Theory, Baoding 071002, China\\
	$^3$ Hebei Research Center of the Basic Discipline for Computational Physics, Baoding 071002, China\\
%	$^4$ College of Science, Hebei Agriculture University, Baoding 071001, China\\
	$^4$ Department of Basic Industrial Education, Hebei Vocational University of Industry and Technology,
	Shijiazhuang 050091, China\\
    $^5$ School of Physics, Harbin Institute of Technology, Harbin 150001, China\\
    $^6$ School of Physical Science and Technology, Northwestern Polytechnical University, Xi'an 710072, China}

\begin{abstract}
We systematically investigate the Dalitz decays of vector heavy quarkonia into $\chi_{QJ}(1P)\ell^+\ell^-$ ($Q=c,b$; $J=0,1,2$; $\ell=e,\mu$) within the instantaneous Bethe--Salpeter framework. The study covers $\psi(2S)$, $\psi(1D)$, $\Upsilon(2S)$, and the so-far unobserved $\Upsilon(1D)$ states. For $\psi(2S)$ electron channels, our predictions are in excellent agreement with BESIII data. We further provide the first relativistic predictions for $\psi(1D)$ decays, with branching ratios for several electron channels reaching the $10^{-5}$ level, which is accessible at current BESIII statistics. For bottomonium, $\Upsilon(2S)\to\chi_{bJ}e^+e^-$ decays yield branching fractions of $\mathcal{O}(10^{-4})$, suggesting potential observability at Belle~II. Muonic channels are also discussed, with kinematic constraints carefully addressed. Our results establish a coherent theoretical basis for future experimental searches for heavy quarkonium Dalitz decays.
\end{abstract}

\maketitle

\section{Introduction}

Charmonium and bottomonium, as bound states of heavy quark--antiquark pairs,
provide an ideal laboratory for studying both perturbative and nonperturbative aspects of QCD~\cite{Chen:2000tv,Brambilla:2010cs}.
Among heavy quarkonia, the electromagnetic Dalitz decays of vector states
$V\to \chi_{QJ}(1P)\ell^+\ell^-$ ($Q=c,b$; $J=0,1,2$; $\ell=e,\mu$),
mediated by a virtual photon converting into a lepton pair,
are particularly sensitive to the non-point-like structure of the bound state and to low-energy strong-interaction dynamics~\cite{Dalitz:1951aj,Zhang:2019xia,Landsberg:1985kjr,Landsberg:1985gaz,Fu:2011yy}.

For charmonium, the $\psi(2S)$ meson (also known as $\psi(3686)$) has been studied extensively \cite{LHCb:2019eaj,BESIII:2017ung,Cao:2016xqo,CLEO:2004cbu,Colangelo:2025yud,Friday:2025gpj,Luth:1975bh,BESIII:2017gcu}.
The first observation of $\psi(2S)\to\chi_{cJ}e^+e^-$ ($J=0,1,2$) was reported in 2017~\cite{BESIII:2017ung},
and related muonic transitions such as $\chi_{cJ}\to\mu^+\mu^- J/\psi$ have also been measured~\cite{BESIII:2019yeu}.
Because $\psi(2S)$ lies below the open-charm threshold, its decay environment is relatively clean and direct theory--experiment comparisons are possible~\cite{LHCb:2019eaj}.
In contrast, the higher vector charmonium $\psi(1D)$, commonly identified with $\psi(3770)$, 
lies just above the $D\bar D$ threshold and has traditionally been studied through its dominant open-charm decays. Nevertheless, BESIII and CLEO have accumulated large $\psi(3770)$ samples and observed several non-$D\bar D$ channels, including radiative transitions $\psi(3770)\to\gamma\chi_{cJ}$ \cite{BESIII:2015cby,BESIII:2015rtt,CLEO:2006nor}.
This makes a systematic study of $\psi(3770)$ Dalitz decays both timely and phenomenologically interesting.

A key kinematic feature distinguishes $\psi(2S)$ from $\psi(3770)$.
For $\psi(2S)$ ($M\simeq 3686$~MeV), only $\psi(2S)\to\chi_{c0}\mu^+\mu^-$ is kinematically allowed,
while the $\chi_{c1}$ and $\chi_{c2}$ muonic channels are forbidden.
By contrast, for $\psi(3770)$ ($M\simeq 3773$~MeV),
all three channels $\psi(3770)\to\chi_{cJ}\mu^+\mu^-$ ($J=0,1,2$) are kinematically allowed,
although the available dilepton invariant-mass range is narrow and the widths are strongly phase-space suppressed.
Thus the muonic Dalitz sector provides a clean illustration of how phase space controls the lepton-flavor structure of charmonium transitions.

Beyond charm, bottomonium offers a complementary testing ground.
The $\Upsilon(2S)$ state is well established and its radiative transitions
$\Upsilon(2S)\to\gamma\chi_{bJ}(1P)$ have been measured at CLEO~\cite{CLEO:1998ukn,CLEO:2004jkt}.
The $\Upsilon(1D)$ state, however, remains experimentally unobserved,
and theoretical studies predict its mass near $10.1$--$10.2$~GeV \cite{Chang:2010kj,Ebert:2011jc,Godfrey:2015dia,Segovia:2016xqb} and discuss its radiative transition properties~\cite{Ebert:2002pp,Radford:2007vd,Godfrey:2015dia,Segovia:2016xqb,Deng:2016ktl,Pei:2022cjy}.
Dalitz decays of bottomonia proceed in close analogy to charmonium,
but with different phase-space boundaries and heavier-quark dynamics,
providing a broader check of any covariant bound-state framework.

Theoretically, 
Dalitz decays are more involved: they contain an off-shell photon, a three-body phase space,
and require careful treatment of the electromagnetic current. The key to the calculation lies in the evaluation of the hadronic transition matrix element, which includes the contributions from nonperturbative QCD.
Many existing studies relate Dalitz rates to radiative ones through transition form factors~\cite{Fu:2011yy,Zhang:2019xia,Colangelo:2025yud}.
In this work we follow another route:
we solve the Bethe-Salpeter (BS) equation in the instantaneous approximation to obtain relativistic Salpeter wave functions,
and evaluate the hadronic transition matrix element as an overlap integral of the initial- and final-state wave functions~\cite{Chang:2006tc,Wang:2005qx,Wang:2022cxy}. This method has been successfully applied to various charmonium decay processes \cite{Fu:2018yxq,Ke:2026gnk,Pei:2024hzv,Li:2023cpl,Li:2022qhg}.
For electron channels, where the lepton mass can be neglected, we impose electromagnetic gauge invariance to stabilize the low-$Q$ region and recover the correct $m_e\to 0$ limit;
for muon channels the larger lepton mass removes the low-$Q$ singularity, so gauge constraints are not required, although phase-space effects become decisive.

In this paper we present a systematic BS study of the Dalitz decays
$\psi(2S)\to\chi_{cJ}(1P)\ell^+\ell^-$,
$\psi(1D)\to\chi_{cJ}(1P)\ell^+\ell^-$,
$\Upsilon(2S)\to\chi_{bJ}(1P)\ell^+\ell^-$,
and $\Upsilon(1D)\to\chi_{bJ}(1P)\ell^+\ell^-$ ($J=0,1,2$; $\ell=e,\mu$).
Our results for $\psi(2S)\to\chi_{cJ}e^+e^-$ agree well with BESIII data.
For $\psi(3770)$ and bottomonium channels we provide, to our knowledge, the first relativistic predictions.
Several channels, such as $\psi(3770)\to\chi_{c0,1}e^+e^-$ and $\Upsilon(2S)\to\chi_{bJ}e^+e^-$, 
have branching fractions at the $10^{-4}$--$10^{-5}$ level,
which are within the reach of BESIII, Belle~II, and future tau--charm facilities.

The paper is organized as follows.
Section~\ref{sec:theory} gives the theoretical formalism, including the invariant amplitude and form-factor decomposition.
Section~\ref{sec:wf} lists the positive-energy BS wave functions for $1^{--}$ and $J^{++}$ states.
Section~\ref{sec:num} presents our results, compared with available data and theoretical predictions.
Section~\ref{sec:summary} summarizes the paper.

\section{Theoretical formalism}
\label{sec:theory}
\begin{figure}[htbp]
		\centering
		\includegraphics[width=\textwidth]{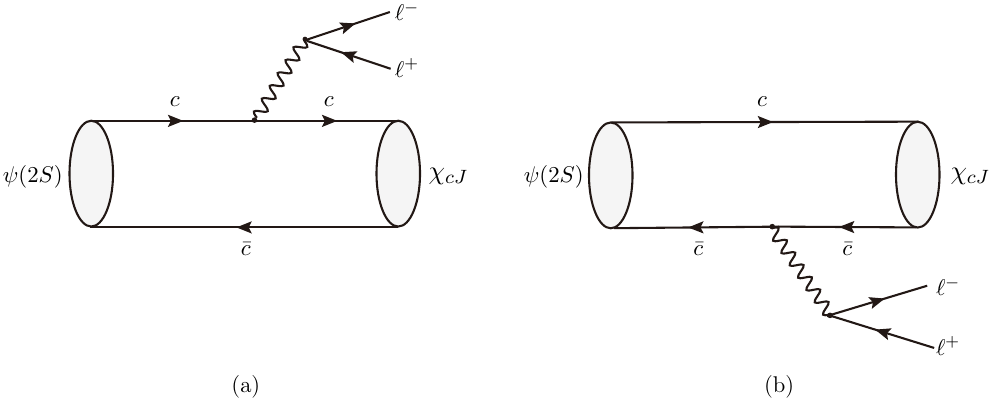}
		\caption{Feynman diagrams for the $\psi(2S) \to \chi_{cJ} \ell^{+}\ell^{-}$ ($\ell = e, \mu$) Dalitz decays.}
		\label{fig:example}
\end{figure}

We take $\psi(2S)\to\chi_{cJ}(1P)\ell^+\ell^-$ ($\ell=e,\mu$) as an example to introduce the Dalitz decay.
Its Feynman diagrams consist of two graphs, as shown in Fig.~\ref{fig:example}.

The corresponding invariant amplitude is expressed as
\begin{eqnarray}
\label{eq1}	\mathcal{M}=\frac{e^2e_{q}}{Q^{2}}\bar{u}_{\ell^-}\gamma_{\mu}v_{\ell^+}\langle{\chi_{cJ}(1P)}|\gamma^{\mu}|{\psi(2S)}\rangle,
\end{eqnarray}
where $e$ is the elementary charge, with $e_{q}=\frac{2}{3}$ for the charm quark and $e_{\bar{q}}=-\frac{2}{3}$ for the anti-charm quark; $Q=\sqrt{(P_{+}+P_{-})^{2}}$ is the invariant mass of the final lepton pair, $P_{+}$ and $P_{-}$ are the four-momenta of the positively and negatively charged leptons; $u$ and $v$ are the spinors for the lepton and anti-lepton; $\langle{\chi_{cJ}}|\gamma^{\mu}|{\psi(2S)}\rangle$ is the hadronic transition matrix element. 

In the initial rest frame, according to the Mandelstam formalism \cite{Mandelstam:1955sd}, we express the hadronic matrix element as an overlap integral of the positive-energy wave functions of the initial and final mesons \cite{Chang:2006tc}:
\begin{flalign}\label{ampli}
	&e_q\langle{\chi_{cJ}(1P)}|\gamma^{\mu}|{\psi(2S)}\rangle\equiv\langle{\chi_{cJ}(1P)}|\gamma^{\mu}|{\psi(2S)}\rangle=
	\nonumber\\
	&\frac{2}{3}\int\frac{d^{3}q_{\bot}}{(2\pi)^{3}} Tr\left\{{\frac{\slashed{P}}{M}\bar{\varphi}^{++}_{f}(q_{\perp}+\frac{1}{2}P_{f\perp})\gamma^{\mu}\varphi^{++}(q_{\perp})-\bar{\varphi}^{++}_{f}(q_{\perp}-\frac{1}{2}P_{f\perp})\frac{\slashed{P}}{M}\varphi^{++}(q_{\perp})\gamma^{\mu}} \right\},
\end{flalign}
where we redefine the transition matrix element to absorb the fractional charge factor; $P$ and $M$ are the initial meson momentum and mass, respectively; $P_f$ is the final meson momentum, and $P_{f\perp}\equiv P_{f}-\frac{P\cdot P_{f}}{M^2}P$ is its perpendicular component to $P$; $q$ is the internal relative momentum of the meson, and for the initial charmonium, we have the following relations, $p_c=\frac{1}{2}P+q$ and $p_{\bar c}=\frac{1}{2}P-q$, with $p_c$ and $p_{\bar c}$ are the momenta of charm quark and anti-charm quark, respectively; in the instantaneous condition, $q\simeq q_{\perp}\equiv q-\frac{P\cdot q}{M^2}P$; $\varphi^{++}(q_{\perp})$ is the positive-energy wave function of the initial meson, and $\varphi^{++}_{f}(q_{f\perp})$ is that of the final meson, with $\bar{\varphi}^{++}=\gamma_0{\varphi}^\dagger\gamma_0$. In the two diagrams of Fig. \ref{fig:example}, the relation between the relative momentum $q_{f\perp}$ of the final meson and the relative momentum $q_{\perp}$ of the initial‑state meson can be obtained from the spectator relation, which we have explicitly given in Eq. (\ref{ampli}), that is, for Fig. \ref{fig:example} (a), $q_{f\perp}=q_{\perp}+\frac{1}{2}P_{f\perp}$, while for Fig. \ref{fig:example} (b), $q_{f\perp}=q_{\perp}-\frac{1}{2}P_{f\perp}$. The explicit wave functions will be given in the next section.

After integration we obtain the form factors:
\begin{flalign}\label{form}
	\langle{\chi_{c0}}|\gamma^{\mu}|{\psi(2S)}\rangle&=t_{1}P^{\mu}(P_f\cdot \varepsilon)+t_{2}P_{f}^{\mu}(P_f\cdot \varepsilon)+t_{3}\varepsilon^{\mu},\nonumber\\
	\langle{\chi_{c1}}|\gamma^{\mu}|{\psi(2S)}\rangle&=s_{1}P^{\mu}\epsilon^{\varepsilon \varepsilon_f P P_f}+s_{2}P_{f}^{\mu}\epsilon^{\varepsilon \varepsilon_f P P_f}+s_{3}\epsilon^{\mu \varepsilon_f P P_f}(P_f\cdot\varepsilon)+s_{4}\epsilon^{\mu \varepsilon \varepsilon_f P_f},\nonumber\\
	\langle{\chi_{c2}}|\gamma^{\mu}|{\psi(2S)}\rangle&=u_{1}P^{\mu}\varepsilon_f(P,P)(P_f\cdot\varepsilon)+u_{2}P_{f}^{\mu}\varepsilon_f(P,P)(P_f\cdot\varepsilon)+u_{3}\varepsilon^{\mu}\varepsilon_f(P,P)\nonumber\\
	&+u_{4}P^{\mu}\varepsilon_f(\varepsilon,P)+u_{5}P_{f}^{\mu}\varepsilon_f(\varepsilon,P)+u_{6}\varepsilon_f(\mu,P)(P_f\cdot\varepsilon)+u_{7}\varepsilon_f(\mu,\varepsilon),
\end{flalign}
where $t_n$, $s_n$, and $u_n$ are form factors; $\epsilon^{\alpha\beta\mu\nu}$ denotes the Levi-Civita symbol, and we have adopted the following shorthand notations: $\epsilon^{\varepsilon \varepsilon_f P P_f}\equiv\epsilon^{\alpha \beta \xi \zeta}\varepsilon_{\alpha} (\varepsilon_f)_{\beta} P_{\xi} (P_f)_{\zeta}$; $\varepsilon^{\mu}$ is the polarization vector of initial meson, $\varepsilon_f^{\mu}$ is the final polarization vector of $\chi_{c1}$, and $\varepsilon_f^{\mu\nu}$ is the second-rank polarization tensor of $\chi_{c2}$, with $\varepsilon_f(P,P)\equiv\varepsilon_f^{\alpha\beta}P_{\alpha}P_{\beta}$.

In the Dalitz decay process of vector charmonium that we consider, the lepton pair originates from a virtual photon rather than a real photon; therefore, this process does not, in principle, need to satisfy gauge invariance (i.e., the Ward identity). However, when the final-state lepton pair is an electron-positron pair, the very small electron mass leads to a divergence-like behavior at small values of the invariant mass $Q$ of the pair, causing the calculated results to become unstable in this region. This is not a genuine divergence problem, but rather a numerical precision issue. Our study shows that imposing the gauge-invariance condition in this small $Q$ region effectively resolves the problem and stabilizes the results, and this also ensures that the results remain correct after the electron mass is neglected. In contrast, when the final-state lepton pair is a muon-antimuon pair, since the muon mass is much larger than the electron mass, this stability issue does not arise, and there is no need to apply the gauge-invariance condition as a constraint.

Therefore, for the electron-positron final-state process, such as $\psi(2S) \to \chi_{cJ} e^+ e^-$, we have imposed the gauge-invariance condition $(P_{\mu}-P_{f\mu})\langle{\chi_{cJ}(1P)}|\gamma^{\mu}|{\psi(2S)}\rangle$=0, when $Q=2m_e\simeq 0$. This implies that the form factors given in Eq. (\ref{form}) are not independent of each other, they satisfy the following relation:
\begin{align}
	\label{eqn4}
	t_3&= (M^2-M\cdot E_f)t_1+(M\cdot E_f-M_f^2)t_2,\nonumber\\
	s_4&=-(M^2-M\cdot E_f)s_1-(M\cdot E_f-M_f^2)s_2,\nonumber\\
	u_7&=-(M^2-M\cdot E_f)u_4-(M\cdot E_f-M_f^2)u_5,\nonumber\\
	u_6&=-(M^2-M\cdot E_f)u_1-(M\cdot E_f-M_f^2)u_2+u_3,
\end{align}
where $M$ is the initial meson mass, $M_f$ the final meson mass, $E_f$ the final meson energy, and $t_n$, $s_n$, $u_n$ are form factors. Then we perform the polarization sum of the squared amplitude:
\begin{align}
	\label{eq5}
	\sum{|M|^2}=\frac{(4\pi)^2\alpha^2}{Q^4}l_{\mu\nu}h^{\mu\nu},
\end{align}
where $l_{\mu\nu}$ is the leptonic tensor:
\begin{align}
	l_{\mu\nu}&=\sum{\bar{u}_{\ell^-}\gamma_{\mu}v_{\ell^+}\bar{v}_{\ell^+}\gamma_{\nu}u_{\ell^-}}\nonumber\\
	&=(-2Q^2+4m^2_{\ell})g_{\mu\nu}+4(P_{+\mu}P_{-\nu}+P_{-\mu}P_{+\nu}),
\end{align}
where $\sum$ denotes the sum over the spins of the lepton and antilepton; and $h^{\mu\nu}$ is the hadronic tensor,
\begin{align}	h^{\mu\nu}&=\sum \langle{\chi_{cJ}(1P)}|\gamma^{\mu}|{\psi(2S)}\rangle \langle{\psi(2S)}|\gamma^{\nu\dagger}|{\chi_{cJ}(1P)}\rangle,
\end{align}
where $\sum$ denotes the sum over the polarizations of the initial and final mesons.

\section{Wave functions}
\label{sec:wf}

In this paper, we perform our calculations using the relativistic wave function, which is obtained by solving the BS equation \cite{Salpeter:1951sz} in the instantaneous approximation \cite{Salpeter:1952ib}. In our method, we do not use the $^{2S+1}L_J$  notation to represent the wave function or a meson; instead, we adopt the $J^{P}$ (or $J^{PC}$) quantum numbers. This is because the conclusion that a meson is a $^{2S+1}L_J$ state is derived only in the nonrelativistic case and does not hold relativistically, whereas the fact that a particle has a definite $J^{PC}$ quantum number is valid in all cases \cite{Wang:2022cxy}. The general wave-function representations for the $1^{--}$ states were first given in Ref. \cite{Wang:2005qx}, and those for the $J^{++}$ ($J=0,1,2$) mesons were provided in Refs. \cite{Wang:2007av,Wang:2007nb,Wang:2009er}. For each of these states, the corresponding full Salpeter equations were solved, and numerical solutions for both the eigenvalues and the radial wave functions were obtained. Interested readers may also refer to Ref. \cite{Wang:2022cxy}, where we present the wave functions for various $J^{PC}$ mesons. Here we do not detail the solving procedure, but only provide the explicit expressions for the positive-energy wave functions based on $J^{PC}$.

The positive-energy wave function for the $1^{--}$ $\psi(2S)$ ($\psi(1D)$) is expressed as \cite{Wang:2005qx}:
\begin{flalign}
	\varphi^{++}_{1^{--}} &= (q_{\bot}\cdot \varepsilon)\left(A_{1}+A_{2}\frac{\slashed{P}}{M}+A_{3}\frac{\slashed{q}_{\bot}}{M}+A_{4}\frac{\slashed{P}\slashed{q}_{\bot}}{M^2}\right)
	\nonumber\\
	&+M \slashed{\varepsilon}\left(A_{5}+A_{6}\frac{\slashed{P}}{M}+A_{7}\frac{\slashed{q}_{\bot}}{M}+A_{8}\frac{\slashed{P}\slashed{q}_{\bot}}{M^2}\right),
\end{flalign}
where the coefficients $A_{n}$ ($n=1\sim8$) are 
\begin{flalign}
	A_1 &= \frac{1}{2M(m_1\omega_2+m_2\omega_1)}[(\omega_1+\omega_2)q_{\perp}^2f_3+(m_1+m_2)q_{\perp}^2f_4+2M^2\omega_2f_5-2M^2m_2f_6], \nonumber\\
	A_2 &= \frac{1}{2M(m_1\omega_2+m_2\omega_1)}[(m_1-m_2)q_{\perp}^2f_3+(\omega_1-\omega_2)q_{\perp}^2f_4-2M^2m_2f_5+2M^2\omega_2f_6], \nonumber\\
	A_3 &= \frac{1}{2}\left[f_3+\frac{m_1+m_2}{\omega_1+\omega_2}f_4-\frac{2M^2}{m_1\omega_2+m_2\omega_1}f_6\right], \nonumber\\
	A_4 &= \frac{1}{2}\left[\frac{\omega_1+\omega_2}{m_1+m_2}f_3+f_4-\frac{2M^2}{m_1\omega_2+m_2\omega_1}f_6\right], \nonumber\\
	A_5 &= \frac{1}{2}\left[f_5-\frac{\omega_1+\omega_2}{m_1+m_2}f_6\right],\nonumber\\
	A_6 &= \frac{1}{2}\left[-\frac{m_1+m_2}{\omega_1+\omega_2}f_5+f_6\right],\nonumber\\
	A_7 &= \frac{M(\omega_1-\omega_2)}{2(m_1\omega_2+m_2\omega_1)}\left[f_5-\frac{\omega_1+\omega_2}{m_1+m_2}f_6\right],\nonumber\\
	A_8 &= \frac{M(m_1+m_2)}{2(m_1\omega_2+m_2\omega_1)}\left[-f_5+\frac{\omega_1+\omega_2}{m_1+m_2}f_6\right],
\end{flalign}
where $m_1$ and $m_2$ are the quark and anti-quark masses; $\omega_1=\sqrt{m_1^2-q^2_{\bot}}$ and $\omega_2=\sqrt{m_2^2-q^2_{\bot}}$ are their energies. As functions of $-q^2_{\bot}$, $f_3$, $f_4$, $f_5$, and $f_6$ are mutually independent radial wave functions, and their numerical solutions as well as the eigenvalue $M$ are obtained by solving the Salpeter equation satisfied by the $1^-$ state \cite{Wang:2005qx}.

For the final $0^{++}$ state $\chi_{c0}$, it's positive-energy wave function can be written as  \cite{Wang:2007av,Wang:2007nb}, 
\begin{flalign}
	\varphi^{++}_{0++} = B_{1}+B_{2}\frac{\slashed{P}_f}{M_{f}}+B_{3}\frac{\slashed{q}_{f\bot}}{M_{f}}+B_{4}\frac{\slashed{P}_f\slashed{q}_{f\bot}}{M_{f}^2},
\end{flalign}
where we have used the subscript $f$ to indicate the final state, and the coefficients $B_i$ ($i=1,2,3,4$)
depend on the independent radial wave functions $g_1(-{q}^2_{f\bot})$ and $g_2(-{q}^2_{f\bot})$, 
\begin{flalign}
	B_1 &= \frac{(\omega_{f1}+\omega_{f2})q_{f\perp}^2}{2(m_{f1}\omega_{f1}+m_{f2}\omega_{f1})}\left[g_1+\frac{m_{f1}+m_{f2}}{\omega_{f1}+\omega_{f2}}g_2\right],\nonumber\\
	B_2 &= \frac{(m_{f1}-m_{f2})q_{f\perp}^2}{2(m_{f1}\omega_{f1}+m_{f2}\omega_{f1})}\left[g_1+\frac{m_{f1}+m_{f2}}{\omega_{f1}+\omega_{f2}}g_2\right],\nonumber\\
	B_3 &= \frac{M_f}{2}\left[g_1+\frac{m_{f1}+m_{f2}}{\omega_{f1}+\omega_{f2}}g_2\right],\nonumber\\
	B_4 &= \frac{M_f}{2}\left[\frac{\omega_{f1}+\omega_{f2}}{m_{f1}+m_{f2}}g_1+g_2\right],
\end{flalign}
where $m_{f1}, m_{f2}$ are the quark masses, $\omega_{f1}=\sqrt{m_{f1}^2-q_{f\bot}^2}$, $\omega_{f2}=\sqrt{m_{f2}^2-q_{f\bot}^2}$. Similarly, the numerical radial wave functions $g_1(-{q}^2_{f\bot})$ and $g_2(-{q}^2_{f\bot})$, as well as the eigenvalue $M_f$ are solutions of full Salpeter equation for $0^{++}$ state \cite{Wang:2007av,Wang:2007nb}.

The positive-energy wave functions for $\chi_{c1}$ and $\chi_{c2}$ (the $1^{++}$ and $2^{++}$ states, respectively) are expressed as \cite{Wang:2007av,Wang:2009er}:
\begin{align}
	\varphi^{++}_{1++}=&i\epsilon_{\mu\nu\alpha\beta}\frac{P_f^\nu}{M_f}q_{f\bot}^{\alpha}\varepsilon^{\beta}_{f}\gamma^{\mu}\left[C_{1}+C_{2}\frac{\slashed{P}_f}{M_{f}}+C_{3}\frac{\slashed{q}_{f\bot}}{M_{f}}+C_{4}\frac{\slashed{P}_f\slashed{q}_{f\bot}}{M_{f}^2}\right],\\
	\varphi^{++}_{2++}=&\epsilon_f(q_{f\bot},q_{f\bot})\left(D_1+\frac{\slashed{P}_f}{M_f}D_2+\frac{\slashed{q}_{f\bot}}{M_f}D_3+\frac{\slashed{P}_f\slashed{q}_{f\bot}}{M_f^2}D_4\right)\nonumber\\&+M_f\epsilon_f(q_{f\bot},\gamma)\left(D_5+\frac{\slashed{P}_f}{M_f}D_6+\frac{\slashed{q}_{f\bot}}{M_f}D_7+\frac{\slashed{P}_f\slashed{q}_{f\bot}}{M_f^2}D_8\right),
\end{align}
with $\epsilon_f(q_{f\bot},q_{f\bot})=(\epsilon_f)_{\mu\nu}q_{f\bot}^{\mu}q_{f\bot}^{\nu}$, $\epsilon_f(q_{f\bot},\gamma)=(\epsilon_f)_{\mu\nu}q_{f\bot}^{\mu}\gamma^\nu$. The coefficients $C_1$--$C_4$ are
\begin{align}
	C_1=\frac{1}{2}\left[h_1+\frac{\omega_{f1}+\omega_{f2}}{m_{f1}+m_{f2}}h_2\right],\nonumber\\
	C_2=\frac{1}{2}\left[\frac{m_{f1}+m_{f2}}{\omega_{f1}+\omega_{f2}}h_1+h_2\right],\nonumber\\
	C_3=\frac{M_f(\omega_{f1}-\omega_{f2})}{m_{f1}\omega_{f2}+m_{f2}\omega_{f1}}C_1,\nonumber\\
	C_4=-\frac{M_f(m_{f1}+m_{f2})}{m_{f1}\omega_{f2}+m_{f2}\omega_{f1}}C_1,
\end{align}
where $h_1(-{q}^2_{f\bot}),h_2(-{q}^2_{f\bot})$ are the independent radial wave function for $1^{++}$ state \cite{Wang:2007av}. The coefficients $D_1$--$D_8$ are
\begin{align}
	D_1 &= \frac{1}{2M(m_{f1}\omega_{f2}+m_{f2}\omega_{f1})}\nonumber\\
	&\times[(\omega_{f1}+\omega_{f2})q_{f\perp}^2k_3+(m_{f1}+m_{f2})q_{f\perp}^2k_4+2M^2\omega_{f2}k_5-2M^2m_{f2}k_6], \nonumber\\
	D_2 &= \frac{1}{2M(m_{f1}\omega_{f2}+m_{f2}\omega_{f1})}\nonumber\\
	&\times[(m_{f1}-m_{f2})q_{f\perp}^2k_3+(\omega_{f1}-\omega_{f2})q_{f\perp}^2k_4-2M^2m_{f2}k_5+2M^2\omega_{f2}k_6], \nonumber\\
	D_3 &= \frac{1}{2}\left[k_3+\frac{m_{f1}+m_{f2}}{\omega_{f1}+\omega_{f2}}k_4-\frac{2M^2}{m_{f1}\omega_{f2}+m_{f2}\omega_{f1}}k_6\right], \nonumber\\
	D_4 &= \frac{1}{2}\left[\frac{\omega_{f1}+\omega_{f2}}{m_{f1}+m_{f2}}k_3+k_4-\frac{2M^2}{m_{f1}\omega_{f2}+m_{f2}\omega_{f1}}k_6\right], \nonumber\\
	D_5 &= \frac{1}{2}\left[k_5-\frac{\omega_{f1}+\omega_{f2}}{m_{f1}+m_{f2}}k_6\right],\nonumber\\
	D_6 &= \frac{1}{2}\left[-\frac{m_{f1}+m_{f2}}{\omega_{f1}+\omega_{f2}}k_5+k_6\right],\nonumber\\
	D_7 &= \frac{M(\omega_{f1}-\omega_{f2})}{2(m_{f1}\omega_{f2}+m_{f2}\omega_{f1})}\left[k_5-\frac{\omega_{f1}+\omega_{f2}}{m_{f1}+m_{f2}}k_6\right],\nonumber\\
	D_8 &= -\frac{M(m_{f1}+m_{f2})}{2(m_{f1}\omega_{f2}+m_{f2}\omega_{f1})}\left[k_5+\frac{\omega_{f1}+\omega_{f2}}{m_{f1}+m_{f2}}k_6\right],
\end{align}
where $k_3$--$k_6$ are independent radial wave functions \cite{Wang:2009er}.

\section{Numerical results}
\label{sec:num}

Details of the radial wave functions adopted in this paper can be found in Ref. \cite{Chang:2010kj}. In that paper, we presented the mass spectrum of charmonium and also plotted the radial wave functions for various $J^{PC}$  states. Therefore, the parameters used in this paper can be found in Ref. \cite{Chang:2010kj}, or in Refs. \cite{Fu:2011tn,Wang:2013lpa}. For example, the charm and bottom quark masses are taken as 1.62 GeV and 4.96 GeV.

\subsection{Charmonium Dalitz decays}
\begin{figure}[htbp]
	\centering
	\subfigure[$\psi(2S) \rightarrow \chi_{c0} e^+ e^-$]{\label{3686_chi_c0_ee}
		\includegraphics[width=0.4\textwidth]{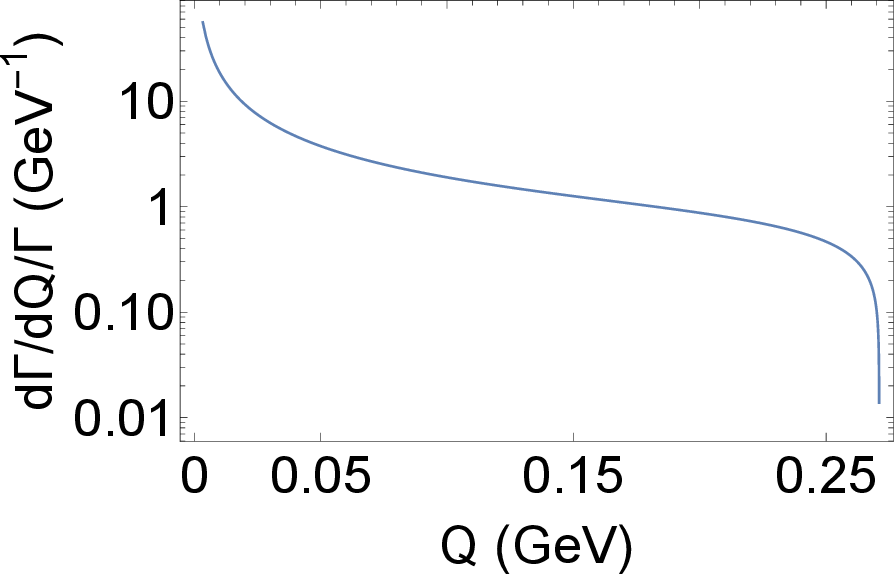}}
	\subfigure[$\psi(2S) \rightarrow \chi_{c0} \mu^+ \mu^-$]{\label{3686_chi_c0_mu}
		\includegraphics[width=0.4\textwidth]{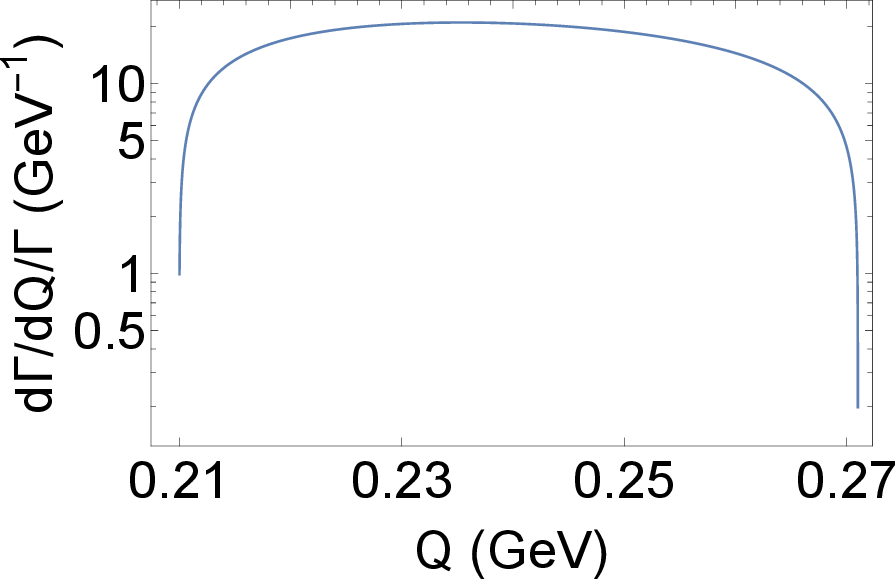}}
	\\
	\subfigure[$\psi(2S) \rightarrow \chi_{c1} e^+ e^-$]{\label{3686_chi_c1_ee}
		\includegraphics[width=0.4\textwidth]{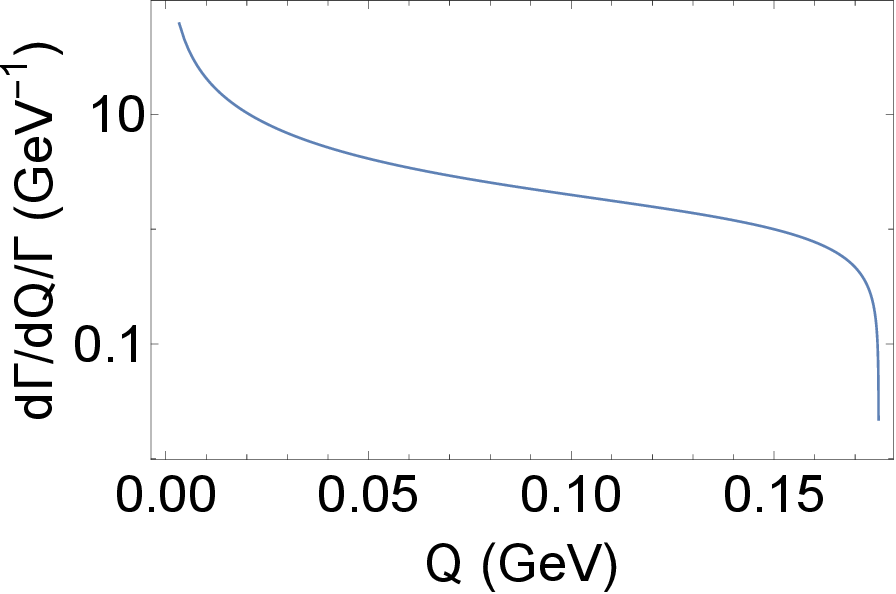}}
	\subfigure[$\psi(2S) \rightarrow \chi_{c2} e^+ e^-$]{\label{3686_chi_c2_ee}
		\includegraphics[width=0.4\textwidth]{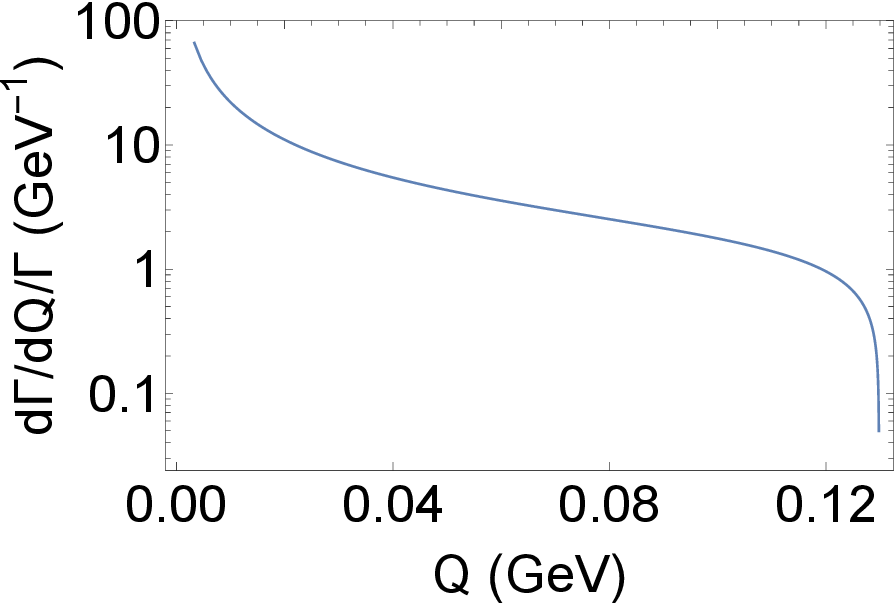}}
	\\
	\subfigure[The spectrum for $\psi(2S) \rightarrow \chi_{c1} e^+ e^-$ by BESIII \cite{BESIII:2017ung}]{\label{chi_c1_e_exp}
		\includegraphics[width=0.4\textwidth]{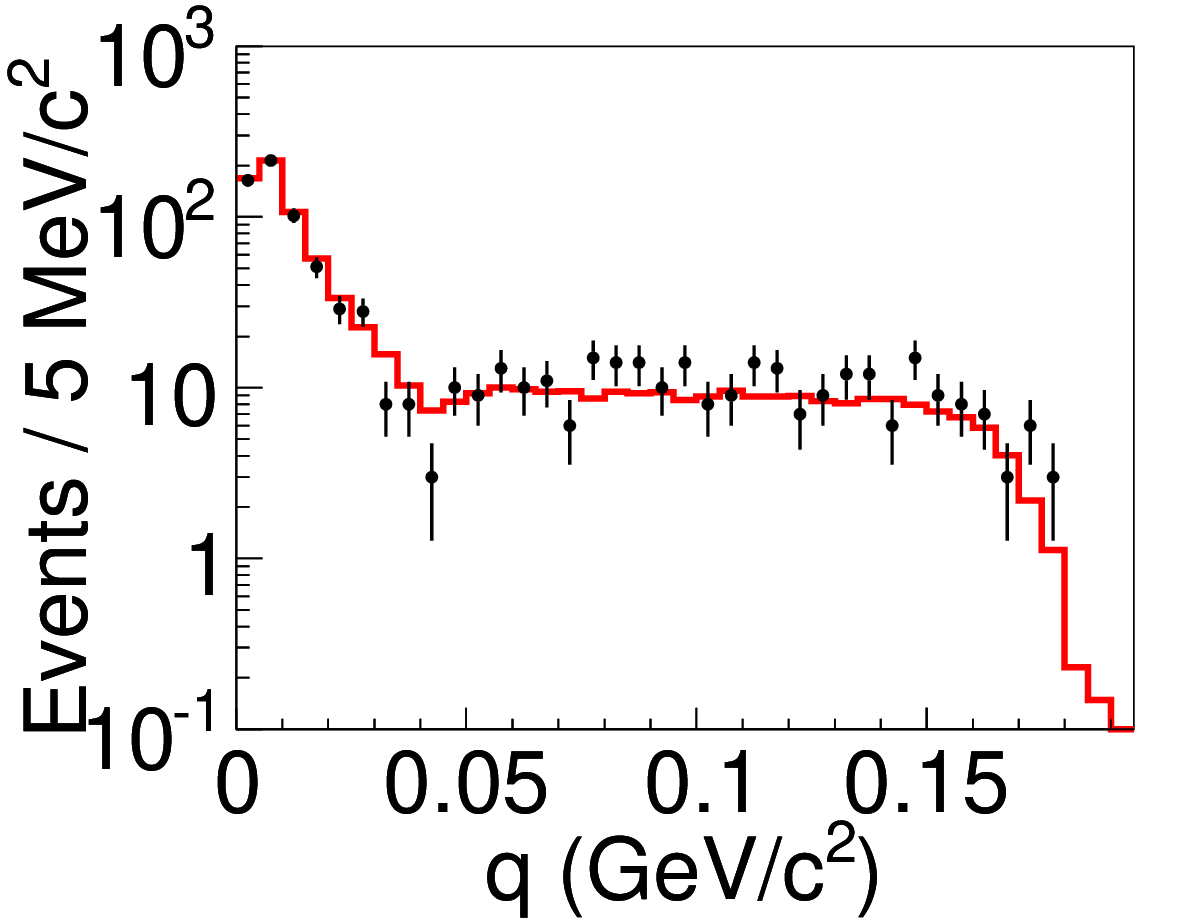}}
	\subfigure[The spectrum for $\psi(2S) \rightarrow \chi_{c2} e^+ e^-$ by BESIII \cite{BESIII:2017ung}]{\label{chi_c2_e_exp}
		\includegraphics[width=0.4\textwidth]{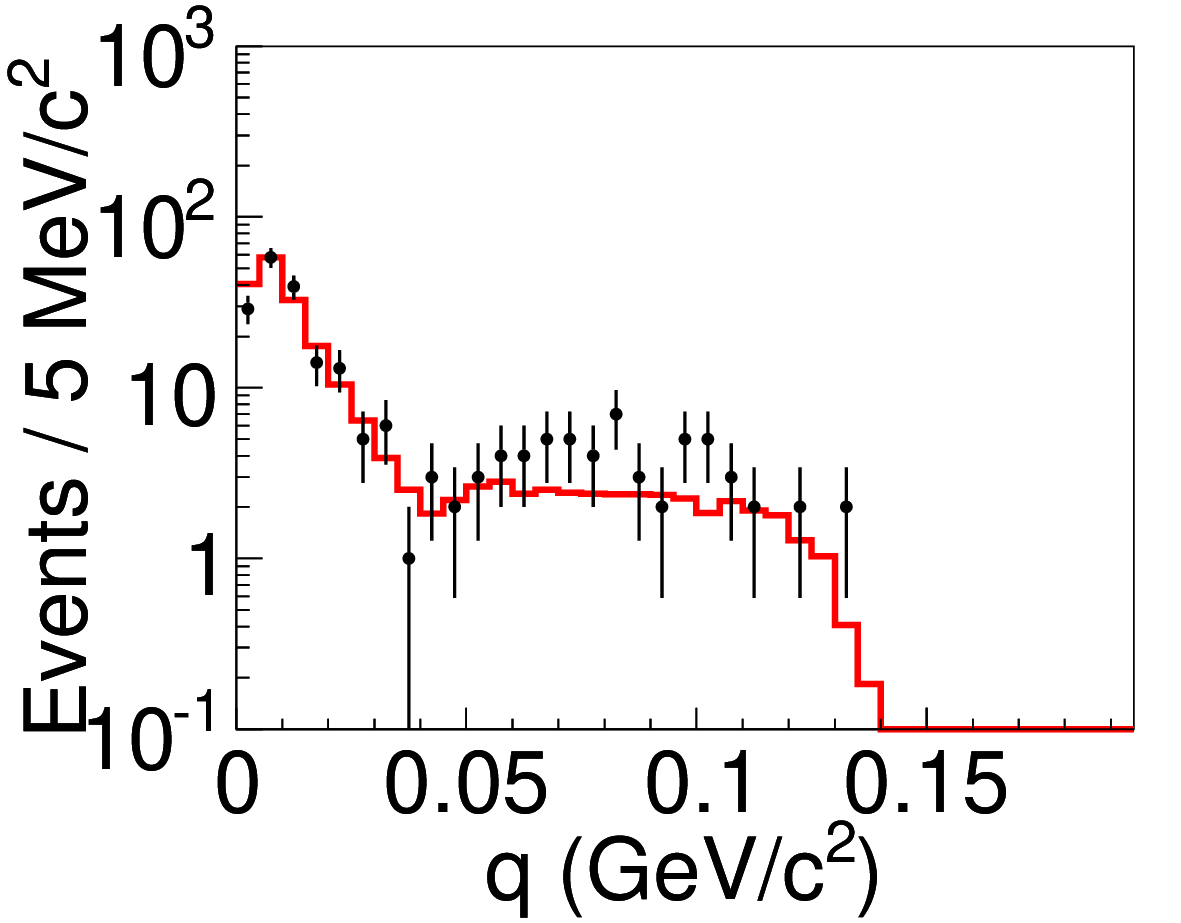}}
	\caption{The invariant mass spectra for the decays $\psi(2S) \to \chi_{cJ} \ell^+ \ell^-$. The experimental data for $\psi(2S) \to \chi_{c1,2} e^+ e^-$ are shown in (e) and (f), where the histograms are for the signal Monte-Carlo (MC) simulation. }
	\label{spc_3686}
	\vspace{-1em}
\end{figure}

We present the calculated results of the normalized differential widths in Figs. \ref{3686_chi_c0_ee}, \ref{3686_chi_c0_mu}, \ref{3686_chi_c1_ee}, and \ref{3686_chi_c2_ee}, where the vertical axis corresponds to $\frac{d\Gamma}{\Gamma dQ}$ and the horizontal axis to $Q$. For comparison, adjacent to Figs. \ref{3686_chi_c1_ee} and \ref{3686_chi_c2_ee}, we also show the experimental data in Figs. \ref{chi_c1_e_exp} and \ref{chi_c2_e_exp}, in which the horizontal axis is denoted as $q$, which is same as our $Q$, i.e., the invariant mass of the final dilepton. Their vertical axis represents the event number. In general, the event number is proportional to the differential width \cite{Colangelo:2025yud}, so a rough comparison between the two can be made. It can be seen that our theoretical curves exhibit the same trend as the experimental data. 

From the figures, it can be seen that for processes with electron-positron pairs in the final state, such as $\psi \to \chi_{cJ} e^+ e^-$, the result curve exhibits a sharp peak in the small $Q$ region. As can be seen from Eq. (\ref{eq1}) or Eq. (\ref{eq5}), the differential width is proportional to $1/Q^4$, while for the electron process, the minimum value of $Q$ is $2m_e$, which leads to the appearance of a sharp peak in the small $Q$ region. This is disastrous for numerical integration, equivalent to neglecting the mass of electron, and the integration result becomes unstable. To address this, we require the corresponding hadronic transition amplitude to satisfy the gauge invariance condition, i.e., Eq. (\ref{eqn4}). On the one hand, this stabilizes the numerical integration results; on the other hand, it also yields correct results in the limit of zero electron mass. We also note that the gauge invariance has also been imposed in other existing theoretical calculations of Refs. \cite{Fu:2011yy,Zhang:2019xia,Colangelo:2025yud}, despite their adoption of entirely different computational methods. 

For the case of muon pairs in the final state, the muon mass is much larger than that of the electron, so no similar peak appears, see the Fig. \ref{3686_chi_c0_mu}, and the transition amplitude does not need to satisfy the gauge invariance condition. 
However, due to the large muon mass, the decays $\psi(2S)\to\chi_{c1}\mu^+\mu^-$ and $\psi(2S)\to\chi_{c2}\mu^+\mu^-$ are phase-space forbidden. By comparing Figs. \ref{3686_chi_c0_ee} and \ref{3686_chi_c0_mu}, one can see that the allowed range of $Q$ in the latter is much smaller than that in the former, which implies that the width obtained by integrating over $Q$ in Fig. \ref{3686_chi_c0_mu} is much smaller than that from Fig. \ref{3686_chi_c0_ee}.

\begin{table}[h]
	\caption{
		Decay widths and branching fractions for the Dalitz decays 
		$\psi(2S) \to \chi_{cJ}\ell^{+}\ell^{-}$ and $\psi(1D) \to \chi_{cJ}\ell^{+}\ell^{-}$ ($J=0,1,2$; $\ell=e,\mu$).
		Theoretical predictions from Ref.~\cite{Colangelo:2025yud} and experimental data from Ref.~\cite{BESIII:2017ung} are also listed for comparison.
	}
	\label{table4}
	\centering
	\begin{tabular}{c c c c c}
		\hline\hline
		Decay mode & $\Gamma$ ($10^{-1}$ keV) & $\mathcal{B}$ & $\mathcal{B}$~\cite{Colangelo:2025yud} & $\mathcal{B}_{\rm BESIII}$~\cite{BESIII:2017ung} \\
		\hline
		$\psi(2S) \to \chi_{c0}e^{+}e^{-}$   & $3.07$              & $1.07\times10^{-3}$ & $(1.11\pm0.03)\times10^{-3}$   & $(1.05\pm0.25)\times10^{-3}$ \\
		$\psi(2S) \to \chi_{c1}e^{+}e^{-}$   & $2.57$              & $8.98\times10^{-4}$ & $(7.6\pm0.2)\times10^{-4}$    & $(8.5\pm0.7)\times10^{-4}$  \\
		$\psi(2S) \to \chi_{c2}e^{+}e^{-}$   & $1.67$              & $5.82\times10^{-4}$ & $(6.6\pm0.2)\times10^{-4}$    & $(6.8\pm0.8)\times10^{-4}$  \\
		\hline
		$\psi(2S) \to \chi_{c0}\mu^{+}\mu^{-}$ & $0.056$ & $1.96\times10^{-5}$ & $(1.51\pm0.04)\times10^{-4}$ & -- \\
		\hline
		$\psi(1D) \to \chi_{c0}e^{+}e^{-}$   & $24.2$      & $8.80\times10^{-5}$ & -- & -- \\
		$\psi(1D) \to \chi_{c1}e^{+}e^{-}$   & $7.26$              & $2.64\times10^{-5}$ & -- & -- \\
		$\psi(1D) \to \chi_{c2}e^{+}e^{-}$   & $0.287$ & $1.04\times10^{-6}$ & -- & -- \\
		\hline
		$\psi(1D) \to \chi_{c0}\mu^{+}\mu^{-}$ & $1.17$              & $4.26\times10^{-6}$ & -- & -- \\
		$\psi(1D) \to \chi_{c1}\mu^{+}\mu^{-}$ & $0.190$ & $6.89\times10^{-7}$ & -- & -- \\
		$\psi(1D) \to \chi_{c2}\mu^{+}\mu^{-}$ & $7.79\times10^{-4}$ & $2.83\times10^{-9}$ & -- & -- \\
		\hline\hline
	\end{tabular}
\end{table}

Table~\ref{table4} lists the decay widths obtained in our work, together with the corresponding branching ratios. For comparison, the theoretical results from Ref.  \cite{Colangelo:2025yud} and the experimental data from Ref. \cite{BESIII:2017ung} are also included in the table. A comparison shows that the branching ratios for $\psi(2S)\to\chi_{c0}e^+e^-$ and $\psi(2S)\to\chi_{c1}e^+e^-$ calculated in our work agree very well with the experimental data, while the branching ratio for $\psi(2S)\to\chi_{c2}e^+e^-$ is slightly lower than the experimental value but still very close. Compared with the theoretical results from Ref. \cite{Colangelo:2025yud}, we find good agreement for $\psi(2S)\to\chi_{c0}e^+e^-$, while channels $\psi(2S)\to\chi_{c1}e^+e^-$ and $\psi(2S)\to\chi_{c2}e^+e^-$ are comparable; however, for $\psi(2S)\to\chi_{c0}\mu^+\mu^-$ the discrepancy is substantial, our result is about one order of magnitude lower than theirs. 

Regarding the decays of the $\psi(3770)$ (i.e., $\psi(1D)$), we have not found any corresponding literature so far. From our results, see Table~\ref{table4}, the branching ratios of $\psi(3770)\to\chi_{cJ}e^+e^-$ and $\psi(3770)\to\chi_{c0}\mu^+\mu^-$ are at the $10^{-5}$ to $10^{-6}$ level, which are relatively small. However, BESIII has accumulated about  $20.3~\text{fb}^{-1}$ of $\psi(3770)$ data, with which its design sensitivity can reach $10^{-6}$--$10^{-7}$ level. So these channels are still within the reach of experimental detection. When the final state consists of two muons, the decays from the $\psi(2S)$ to $\chi_{c1}$ and $\chi_{c2}$ are phase-space forbidden. In contrast, the phase space for the $\psi(3770) \to \chi_{cJ}\mu^+\mu^-$ decays is small, which makes the decay widths much smaller than those of the corresponding electron channels, especially for the case of $\chi_{c2}$, whose branching ratio is at the  level of $10^{-9}$.

\begin{figure}[htbp]
	\centering
	\subfigure[$\psi(1D) \to \chi_{c0} e^+ e^-$]{\label{3770_chi_c0_ee}
		\includegraphics[width=0.4\textwidth]{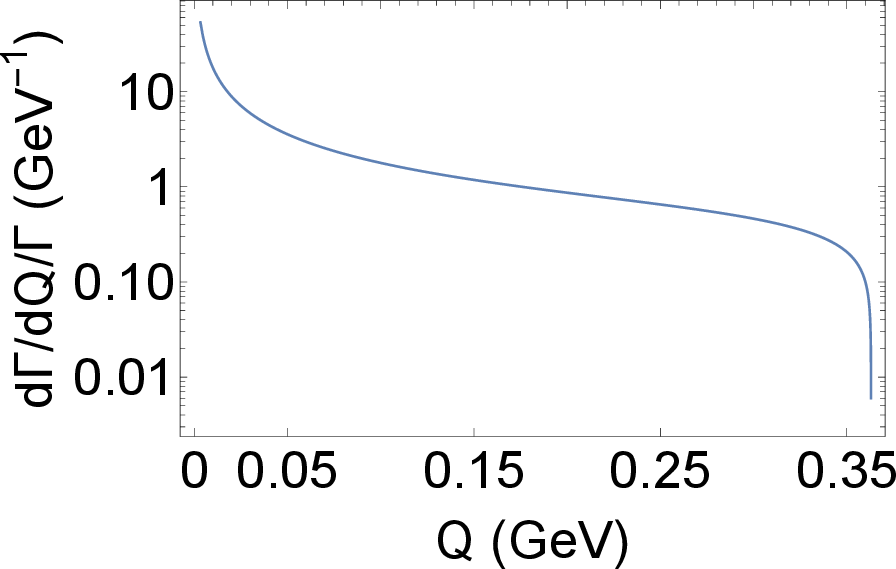}}
	\subfigure[$\psi(1D) \rightarrow \chi_{c0} \mu^+ \mu^-$]{\label{3770_chi_c0_mu}
		\includegraphics[width=0.4\textwidth]{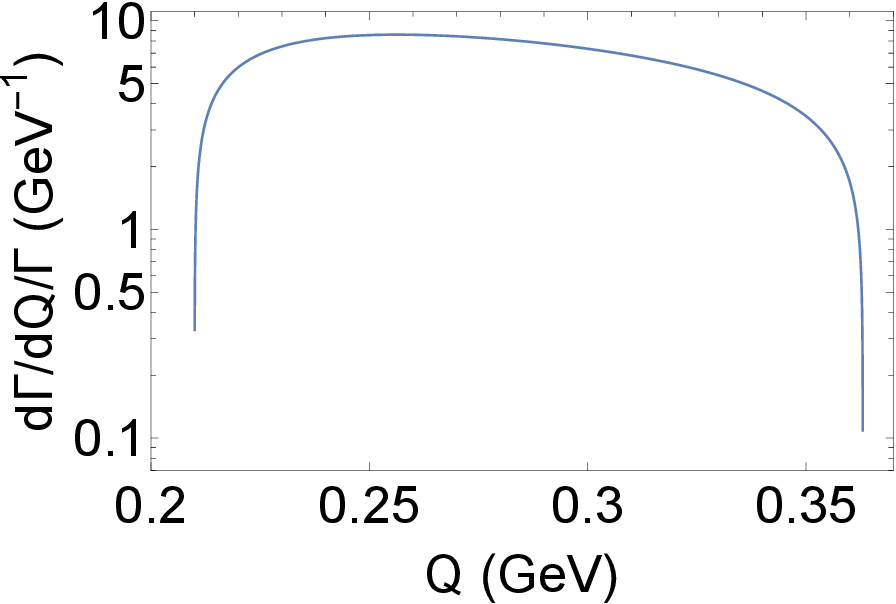}}
	\\
	\subfigure[$\psi(1D) \rightarrow \chi_{c1} e^+ e^-$]{\label{3770_chi_c1_ee}
		\includegraphics[width=0.4\textwidth]{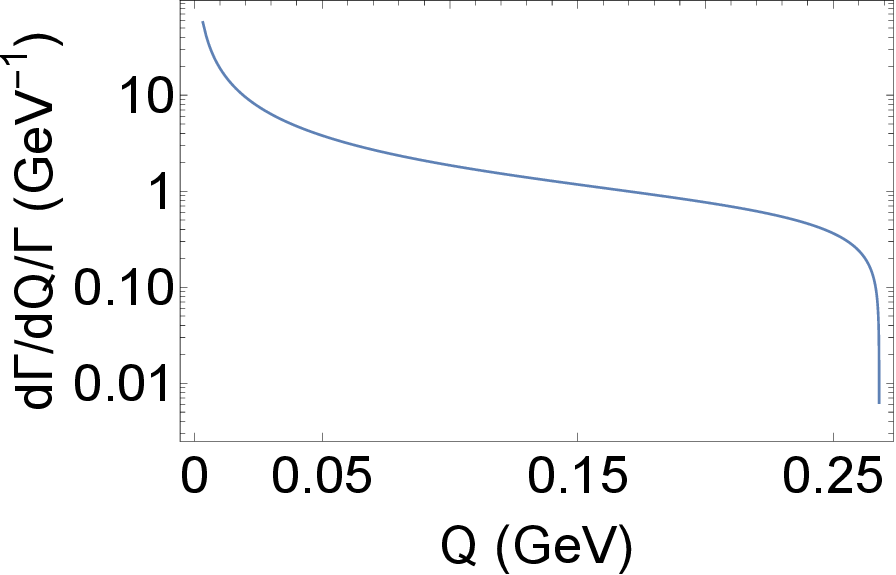}}
	\subfigure[$\psi(1D) \rightarrow \chi_{c1} \mu^+ \mu^-$]{\label{377x0_chi_c1_mu}
		\includegraphics[width=0.4\textwidth]{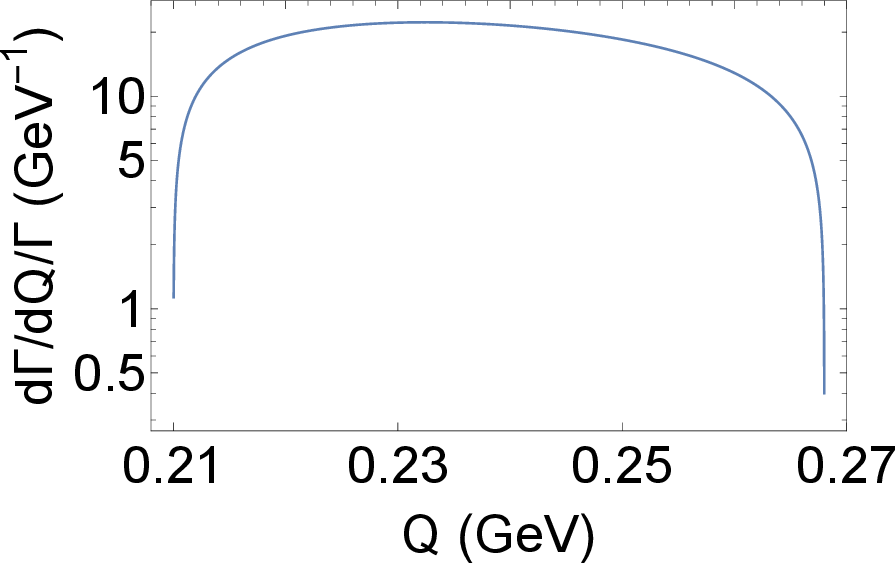}}
	\\
	\subfigure[$\psi(1D) \rightarrow \chi_{c2} e^+ e^-$]{\label{3770_chi_c2_ee}
		\includegraphics[width=0.4\textwidth]{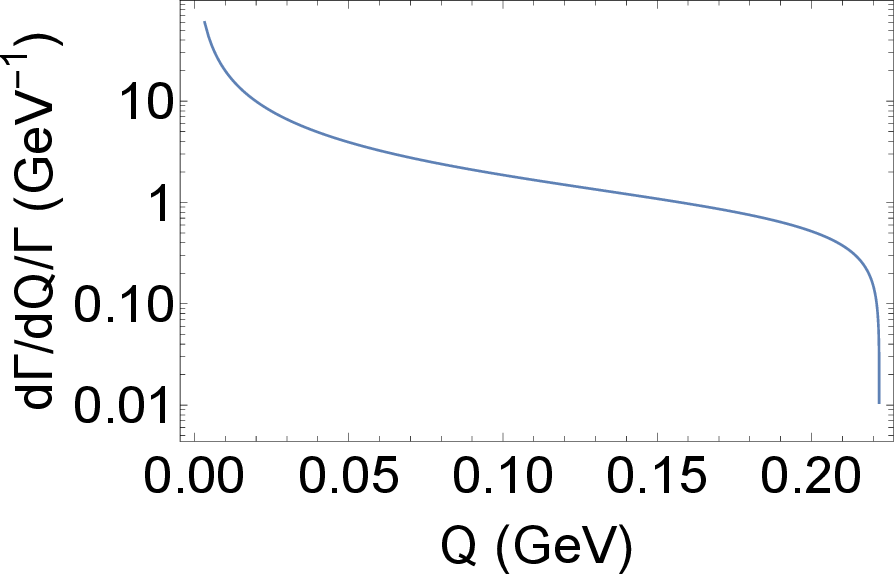}}
	\subfigure[$\psi(1D) \rightarrow \chi_{c2} \mu^+ \mu^-$]{\label{3770_chi_c2_mu}
		\includegraphics[width=0.4\textwidth]{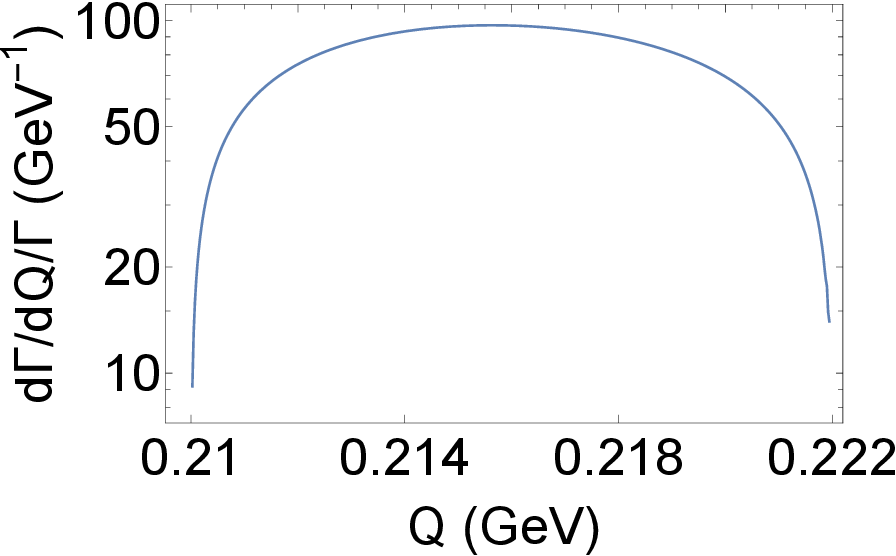}}
	\caption{The invariant mass spectra for the decays $\psi(1D) \to \chi_{cJ} \ell^+ \ell^-$. }
	\label{spc_3770}
	\vspace{-1em}
\end{figure}

In Fig. \ref{spc_3770}, we present the curves of the differential widths for $\psi(1D) \to \chi_{cJ} \ell^+\ell^-$ as functions of the dilepton invariant mass. As can be seen from the figures, similar to the $\psi(2S)$ case, a sharp peak also appears at very small $Q$ range for the $\psi(1D) \to \chi_{cJ} e^+e^-$ process, while this structure is absent in the $\psi(1D) \to \chi_{cJ} \mu^+\mu^-$ plots.

\begin{table}[ht]
\begin{center}
	\caption{
		Ratios of Dalitz to radiative decay branching fractions,
		$\mathcal{B}(M' \to M\ell^{+}\ell^{-})/\mathcal{B}(M' \to M\gamma)$ in unit of $10^{-3}$,
		and lepton-flavor ratios,
		$\mathcal{B}(M' \to M\mu^{+}\mu^{-})/\mathcal{B}(M' \to M e^{+}e^{-})$  in unit of $10^{-2}$,
		for charmonium systems.
	}
	\label{table2}
{
		\begin{tabular}{cc cccc }\hline\hline
Ratio&This work&Ref.\cite{Zhang:2019xia}&BESIII\cite{BESIII:2017ung}&~Ratio~&~~This work
			   \\\hline
~~$\frac{\mathcal{B}(\psi(2S) \rightarrow \chi_{c0}e^{+}e^{-})}{\mathcal{B}(\psi(2S) \rightarrow \chi_{c0} \gamma)}$&7.69&7.95&$9.4\pm1.9\pm0.6$~~&~~~~$\frac{\mathcal{B}(\psi(2S) \rightarrow \chi_{c0}\mu^{+}\mu^{-})}{\mathcal{B}(\psi(2S) \rightarrow \chi_{c0} e^{+}e^{-})}$~~&~ $1.83$
			   \\ 
~~$\frac{\mathcal{B}(\psi(2S) \rightarrow \chi_{c1}e^{+}e^{-})}{\mathcal{B}(\psi(2S) \rightarrow \chi_{c1} \gamma)}$&7.22& 7.28&$8.3\pm0.3\pm0.4$~~& --& --
			   \\
~~$\frac{\mathcal{B}(\psi(2S) \rightarrow \chi_{c2}e^{+}e^{-})}{\mathcal{B}(\psi(2S) \rightarrow \chi_{c2} \gamma)}$&6.82&6.81 &~~$6.6\pm0.5\pm0.4$~~~~& --& --
			   \\\hline
~~$\frac{\mathcal{B}(\psi(1D) \rightarrow \chi_{c0}e^{+}e^{-})}{\mathcal{B}(\psi(1D) \rightarrow \chi_{c0} \gamma)}$&8.34& -- & --&~~~~$\frac{\mathcal{B}(\psi(1D) \rightarrow \chi_{c0}\mu^{+}\mu^{-})}{\mathcal{B}(\psi(1D) \rightarrow \chi_{c0} e^{+}e^{-})}$&~~ $4.84$
			   \\	
~~$\frac{\mathcal{B}(\psi(1D) \rightarrow \chi_{c1} e^{+}e^{-})}{\mathcal{B}(\psi(1D) \rightarrow \chi_{c1})\gamma}$&8.00& --& --&~~~~$\frac{\mathcal{B}(\psi(1D) \rightarrow \chi_{c1}\mu^{+}\mu^{-})}{\mathcal{B}(\psi(1D) \rightarrow \chi_{c1} e^{+}e^{-})}$&~~ $2.61$
			   \\
~~$\frac{\mathcal{B}(\psi(1D) \rightarrow \chi_{c2}e^{+}e^{-})}{\mathcal{B}(\psi(1D) \rightarrow \chi_{c2} \gamma)}$&8.20&  --& --&~~~~$\frac{\mathcal{B}(\psi(1D) \rightarrow \chi_{c2}\mu^{+}\mu^{-})}{\mathcal{B}(\psi(1D) \rightarrow \chi_{c2} e^{+}e^{-})}$&~~ $0.272$
			  \\ \hline\hline
		\end{tabular}}
\end{center}
\end{table}

Table \ref{table2} presents the ratio $\frac{\mathcal{B}(M' \rightarrow M\ell^{+}\ell^{-})}{\mathcal{B}(M' \rightarrow M \gamma)}$, where the branching ratios for the radiative electromagnetic decays $\mathcal{B}(M' \rightarrow M \gamma)$ are taken from our previous work \cite{Pei:2022cjy}. The theoretical results \cite{Zhang:2019xia} and the BESIII experimental data \cite{BESIII:2017ung} are also shown in Table \ref{table2} for comparison. Our results are in excellent agreement with those of Ref. \cite{Zhang:2019xia} and also show good agreement with the experimental values. We find that this ratio falls within a relatively narrow range: $(6.82-7.69)\times 10^{-3}$ for $\frac{\mathcal{B}(\psi(2S) \rightarrow \chi_{cJ}e^{+}e^{-})}{\mathcal{B}(\psi(2S) \rightarrow \chi_{cJ}\gamma)}$ and an even narrower range of $(8.0-8.34)\times 10^{-3}$ for $\frac{\mathcal{B}(\psi(1D) \rightarrow \chi_{cJ}e^{+}e^{-})}{\mathcal{B}(\psi(1D) \rightarrow \chi_{cJ}\gamma)}$. This reflects the additional factor of $\alpha$ in Dalitz decays compared to radiative electromagnetic decays.

In addition, Table \ref{table2} also gives the ratio $\frac{\mathcal{B}(M' \rightarrow M\mu^{+}\mu^{-})}{\mathcal{B}(M' \rightarrow M e^{+}e^{-})}$. Our theoretical results show that the ratios $\frac{\mathcal{B}(\psi(2S) \rightarrow \chi_{c0}\mu^{+}\mu^{-})}{\mathcal{B}(\psi(2S) \rightarrow \chi_{c0} e^{+}e^{-})}$ and $\frac{\mathcal{B}(\psi(1D) \rightarrow \chi_{cJ}\mu^{+}\mu^{-})}{\mathcal{B}(\psi(1D) \rightarrow \chi_{cJ} e^{+}e^{-})}$ typically lies at the $\mathcal{O}(10^{-2}-10^{-3})$ level.

\subsection{Bottomonium Dalitz decays}

\begin{figure}[htbp]
	\centering
	\subfigure[$\Upsilon(2S) \to \chi_{b0} e^+ e^-$]{\label{kUpsilon2S_chib0_e}
		\includegraphics[width=0.4\textwidth]{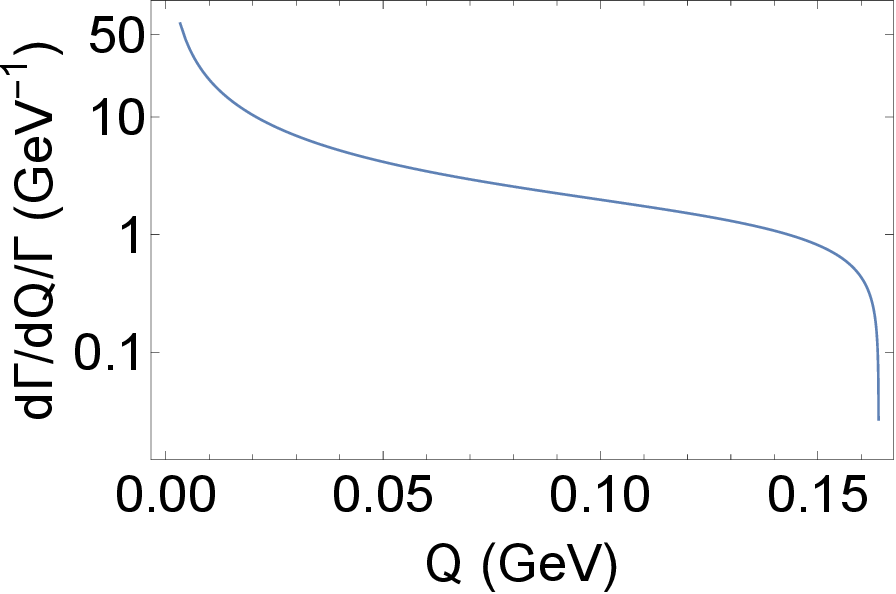}}
	\subfigure[$\Upsilon(2S) \rightarrow \chi_{b1} e^+ e^-$]{\label{kUpsilon2S_chib1_e}
		\includegraphics[width=0.4\textwidth]{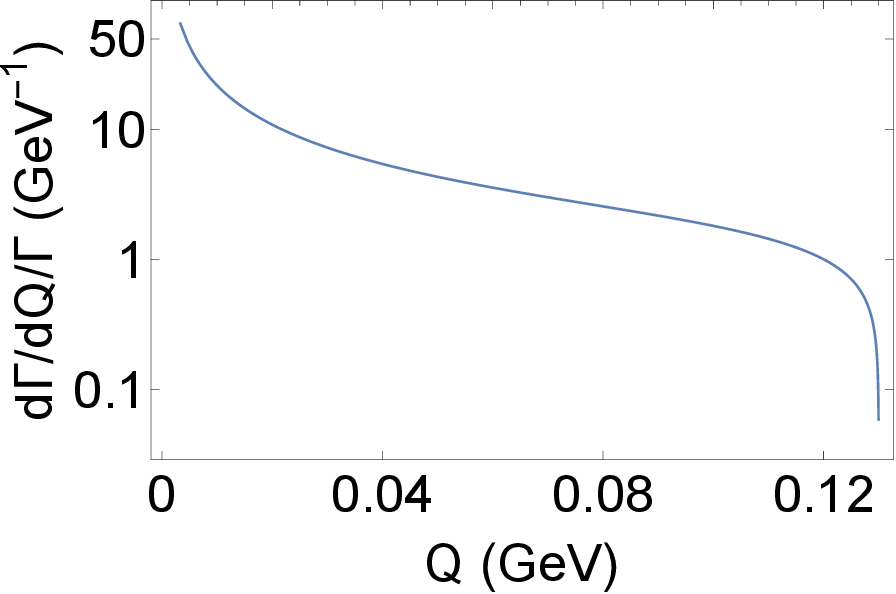}}
	\\
	\subfigure[$\Upsilon(2S) \rightarrow \chi_{b2} e^+ e^-$]{\label{kUpsilon2S_chib2_e}
		\includegraphics[width=0.4\textwidth]{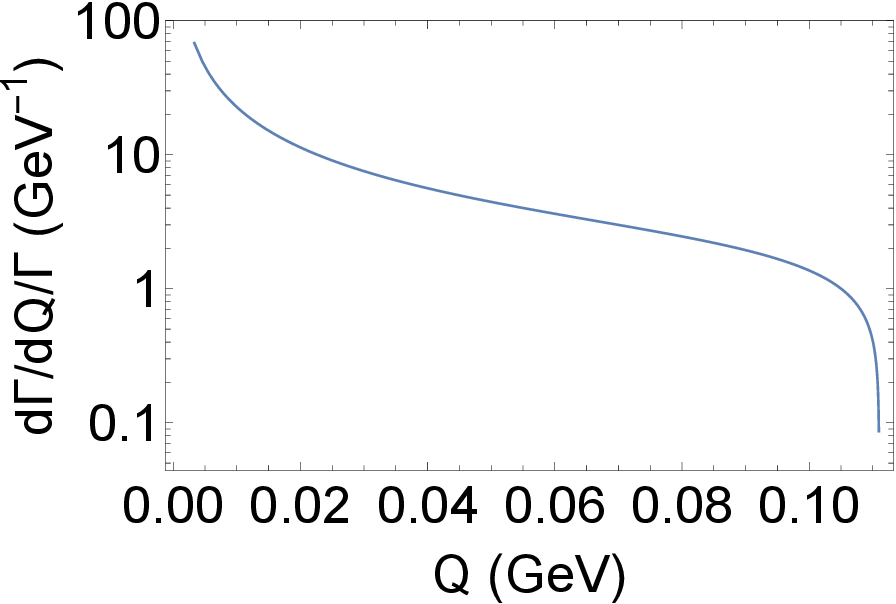}}
	\caption{The invariant mass spectra for the decays $\Upsilon(2S) \to \chi_{bJ} e^+ e^-$. }
	\label{Upsilon2S}
	\vspace{-1em}
\end{figure}

For the $\Upsilon(2S)$ Dalitz decays, only the dielectron channel is kinematically allowed. We plot the differential width as a function of the dilepton invariant mass spectrum in Fig. \ref{Upsilon2S}. The overall shape is similar to that of the charmonium cases, and we have also imposed the gauge invariance condition here, which we will not elaborate on further. 

\begin{figure}[htbp]
	\centering
	\subfigure[$\Upsilon(1D) \to \chi_{b0} e^+ e^-$]{\label{kUpsilon1D_chib0_e}
		\includegraphics[width=0.4\textwidth]{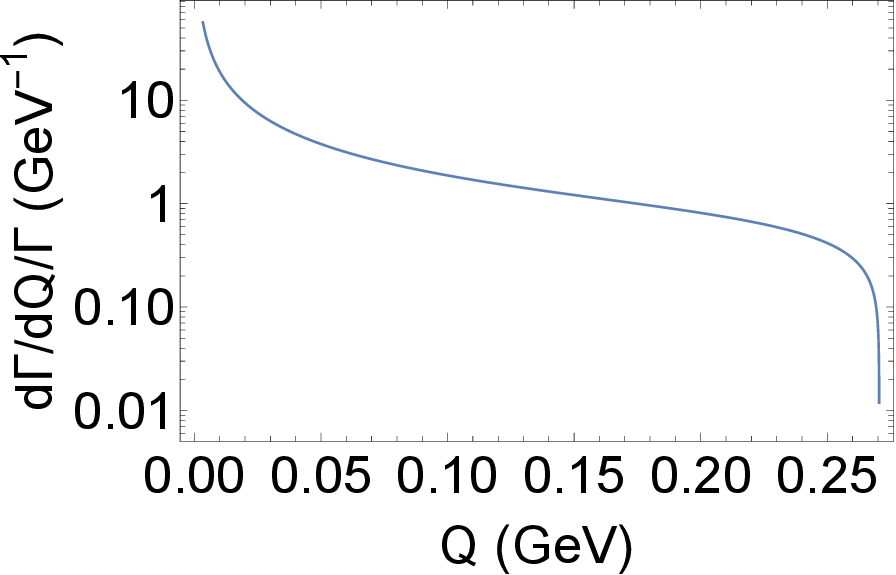}}
	\subfigure[$\Upsilon(1D) \rightarrow \chi_{b0} \mu^+ \mu^-$]{\label{kUpsilon1D_chib0_mu}
		\includegraphics[width=0.4\textwidth]{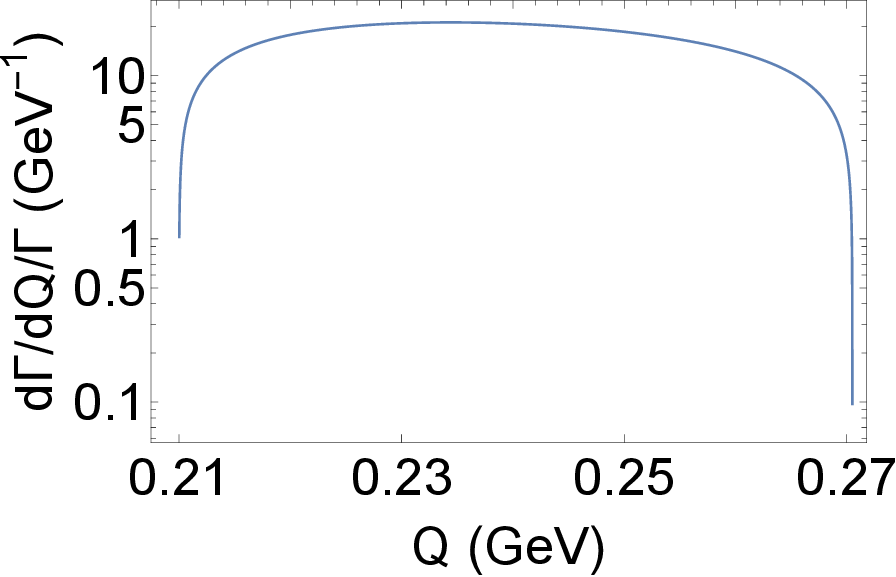}}
	\\
	\subfigure[$\Upsilon(1D) \rightarrow \chi_{b1} e^+ e^-$]{\label{kUpsilon1D_chib1_e}
		\includegraphics[width=0.4\textwidth]{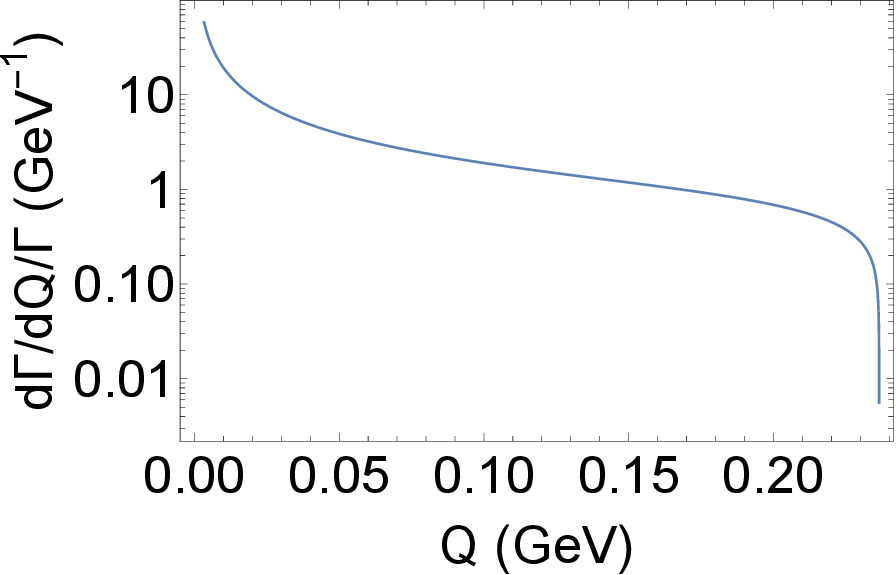}}
	\subfigure[$\Upsilon(1D) \rightarrow \chi_{b1} \mu^+ \mu^-$]{\label{kUpsilon1D_chib1_mu}
		\includegraphics[width=0.4\textwidth]{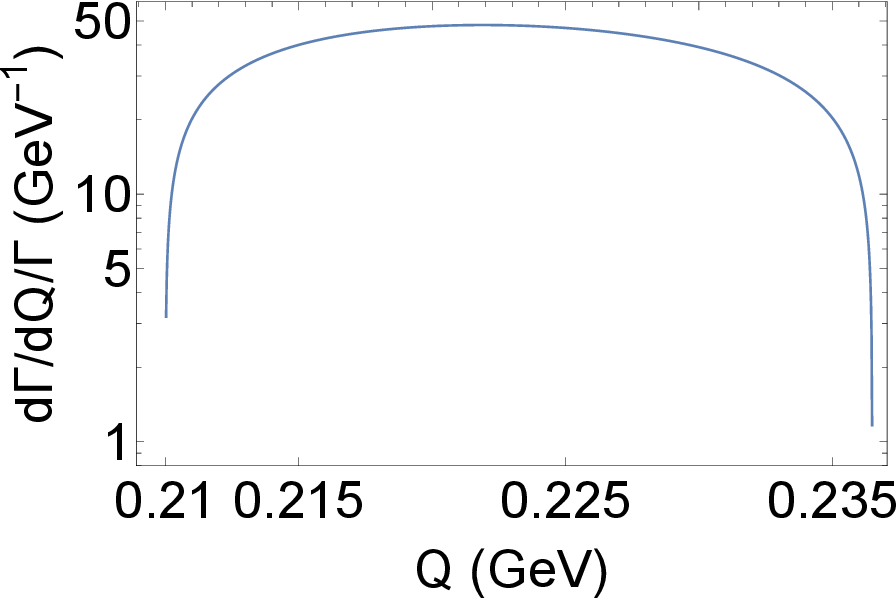}}
	\\
	\subfigure[$\Upsilon(1D) \rightarrow \chi_{b2} e^+ e^-$]{\label{kUpsilon1D_chib2_e}
		\includegraphics[width=0.4\textwidth]{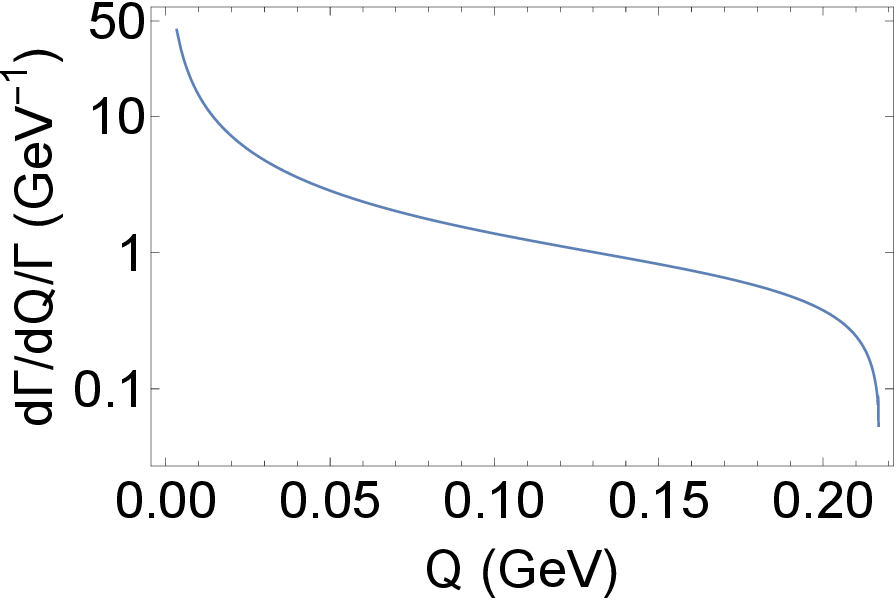}}
	\subfigure[$\Upsilon(1D) \rightarrow \chi_{b2} \mu^+ \mu^-$]{\label{kUpsilon1D_chib2_mu}
		\includegraphics[width=0.4\textwidth]{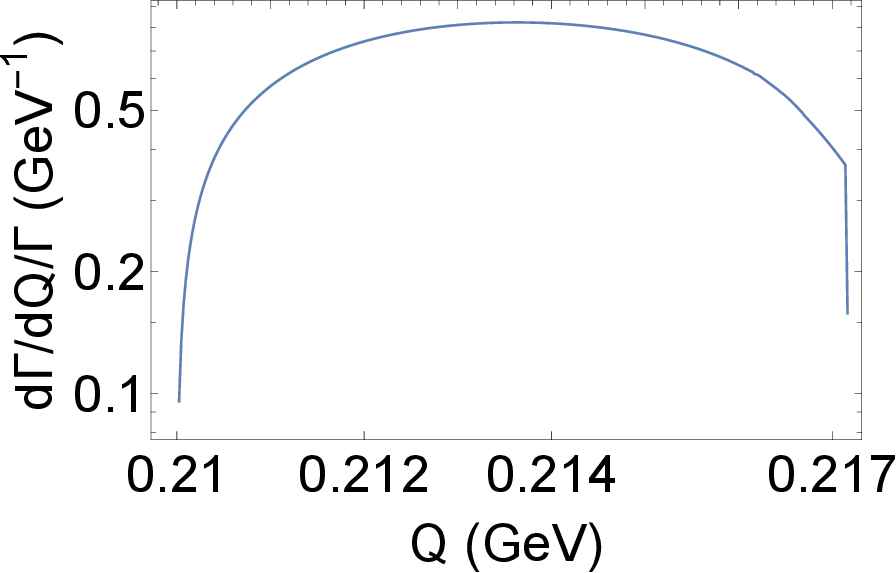}}
	\caption{The invariant mass spectra for the decays $\Upsilon(1D) \to \chi_{bJ} \ell^+ \ell^-$. }
	\label{Upsilon1D}
	\vspace{-1em}
\end{figure}

The $\Upsilon(1D)$ state has not been experimentally discovered so far, and its mass is taken from the theoretical value calculated in our previous work, $m = 10129.5~\text{MeV}$ \cite{Chang:2010kj}. At this mass, the Dalitz decays of $\Upsilon(1D)$ also include the dimuon channel. Their differential width spectra are plotted in Fig. \ref{Upsilon1D}. Similarly, the dimuon channel does not require the imposition of the gauge invariance condition; however, one can see that the allowed range of the invariant mass $Q$ is very small, so the corresponding decay widths will be very small.

\begin{table}[ht]
	\centering
	\caption{
		Decay widths and branching fractions for the Dalitz decays
		$\Upsilon(2S) \to \chi_{bJ}(1P)\ell^{+}\ell^{-}$
		and $\Upsilon(1D) \to \chi_{bJ}(1P)\ell^{+}\ell^{-}$
		($J=0,1,2$; $\ell=e,\mu$).
	}
	\label{tablebb}
	\begin{tabular}{c c c}
		\hline\hline
		Decay mode & $\Gamma$ (GeV) & $\mathcal{B}$ \\
		\hline
		$\Upsilon(2S) \to \chi_{b0}e^{+}e^{-}$ & $8.25\times10^{-9}$  & $2.58\times10^{-4}$ \\
		$\Upsilon(2S) \to \chi_{b1}e^{+}e^{-}$ & $1.20\times10^{-8}$  & $3.75\times10^{-4}$ \\
		$\Upsilon(2S) \to \chi_{b2}e^{+}e^{-}$ & $1.24\times10^{-8}$  & $3.88\times10^{-4}$ \\
		\hline
		$\Upsilon(1D) \to \chi_{b0}e^{+}e^{-}$ & $1.22\times10^{-7}$  & -- \\
		$\Upsilon(1D) \to \chi_{b1}e^{+}e^{-}$ & $6.13\times10^{-8}$  & -- \\
		$\Upsilon(1D) \to \chi_{b2}e^{+}e^{-}$ & $4.23\times10^{-9}$  & -- \\
		\hline
		$\Upsilon(1D) \to \chi_{b0}\mu^{+}\mu^{-}$ & $2.13\times10^{-9}$  & -- \\
		$\Upsilon(1D) \to \chi_{b1}\mu^{+}\mu^{-}$ & $3.57\times10^{-10}$ & -- \\
		$\Upsilon(1D) \to \chi_{b2}\mu^{+}\mu^{-}$ & $4.16\times10^{-10}$ & -- \\
		\hline\hline
	\end{tabular}
\end{table}
	
The decay widths and the corresponding branching ratios for $\Upsilon(2S)$, as well as the decay widths for $\Upsilon(1D)$, are all listed in Table \ref{tablebb}. As can be seen from the table, the decay widths of $\Upsilon(2S)$ are relatively small, ranging from a few eV to about ten eV. Despite this smallness, since the total width of $\Upsilon(2S)$ is $32~\text{keV}$, the corresponding branching ratios are at the $10^{-4}$ level, which still lies within the reach of experimental detection. For the $\Upsilon(1D)$ Dalitz decays, the widths vary significantly, from the largest one, $\Upsilon(1D) \to \chi_{b0} e^+e^-$ ($122~\text{eV}$), down to the very small $\Upsilon(1D) \to \chi_{b2} \mu^+\mu^-$ ($0.416~\text{eV}$).

\begin{table}[ht]
	\centering
	\caption{
		Ratios of Dalitz to radiative decay branching fractions,
		$\mathcal{B}(M' \to M\ell^{+}\ell^{-})/\mathcal{B}(M' \to M\gamma)$,
		and lepton-flavor ratios,
		$\mathcal{B}(M' \to M\mu^{+}\mu^{-})/\mathcal{B}(M' \to M e^{+}e^{-})$,
		for bottomonium systems.	}
	\label{table41}
	\begin{tabular}{c c c c}
		\hline\hline
		Ratio & Value ($10^{-3}$) & Ratio & Value ($10^{-2}$) \\
		\hline
		$\frac{\mathcal{B}(\Upsilon(2S) \to \chi_{b0}e^{+}e^{-})}{\mathcal{B}(\Upsilon(2S) \to \chi_{b0} \gamma)}$ & 7.30 & -- & -- \\
		$\frac{\mathcal{B}(\Upsilon(2S) \to \chi_{b1}e^{+}e^{-})}{\mathcal{B}(\Upsilon(2S) \to \chi_{b1} \gamma)}$ & 6.67 & -- & -- \\
		$\frac{\mathcal{B}(\Upsilon(2S) \to \chi_{b2}e^{+}e^{-})}{\mathcal{B}(\Upsilon(2S) \to \chi_{b2} \gamma)}$ & 6.78 & -- & -- \\
		\hline
		$\frac{\mathcal{B}(\Upsilon(1D) \to \chi_{b0}e^{+}e^{-})}{\mathcal{B}(\Upsilon(1D) \to \chi_{b0} \gamma)}$ & 7.87 & $\frac{\mathcal{B}(\Upsilon(1D) \to \chi_{b0}\mu^{+}\mu^{-})}{\mathcal{B}(\Upsilon(1D) \to \chi_{b0} e^{+}e^{-})}$ & 1.75 \\
		$\frac{\mathcal{B}(\Upsilon(1D) \to \chi_{b1} e^{+}e^{-})}{\mathcal{B}(\Upsilon(1D) \to \chi_{b1}\gamma)}$ & 7.72 & $\frac{\mathcal{B}(\Upsilon(1D) \to \chi_{b1}\mu^{+}\mu^{-})}{\mathcal{B}(\Upsilon(1D) \to \chi_{b1} e^{+}e^{-})}$ & 0.582 \\
		$\frac{\mathcal{B}(\Upsilon(1D) \to \chi_{b2}e^{+}e^{-})}{\mathcal{B}(\Upsilon(1D) \to \chi_{b2} \gamma)}$ & 10.2 & $\frac{\mathcal{B}(\Upsilon(1D) \to \chi_{b2}\mu^{+}\mu^{-})}{\mathcal{B}(\Upsilon(1D) \to \chi_{b2} e^{+}e^{-})}$ & 9.83 \\
		\hline\hline
	\end{tabular}
\end{table}

In table \ref{table41}, the ratios $\frac{\mathcal{B}(M' \rightarrow M\ell^{+}\ell^{-})}{\mathcal{B}(M' \rightarrow M \gamma)}$ and $\frac{\mathcal{B}(M' \rightarrow M\mu^{+}\mu^{-})}{\mathcal{B}(M' \rightarrow M e^{+}e^{-})}$ are shown, where the results of the radiative decays $\mathcal{B}(M' \rightarrow M \gamma)$ are also taken from Ref. \cite{Pei:2022cjy}. The majority of the ratios $\frac{\mathcal{B}(M' \rightarrow M\ell^{+}\ell^{-})}{\mathcal{B}(M' \rightarrow M \gamma)}$ fall in a narrow range of $6.67$-$7.87$, except for $\frac{\mathcal{B}(\Upsilon(1D) \rightarrow \chi_{b2}e^{+}e^{-})}{\mathcal{B}(\Upsilon(1D) \rightarrow \chi_{b2}\gamma)}=10.2$. While the magnitude of $\frac{\mathcal{B}(\Upsilon(1D) \rightarrow \chi_{bJ}\mu^{+}\mu^{-})}{\mathcal{B}(\Upsilon(1D) \rightarrow \chi_{bJ} e^{+}e^{-})}$ is $\mathcal{O}(10^{-2}$-$10^{-3})$, its behavior differs markedly from the charmonium case.

\section{Summary}
\label{sec:summary}

In this work, we have presented a systematic investigation of the Dalitz decays 
$V \to \chi_{QJ}(1P) \ell^+\ell^-$ ($V = \psi(2S), \psi(1D), \Upsilon(2S), \Upsilon(1D)$; $Q=c,b$; $J=0,1,2$; $\ell=e,\mu$) within the instantaneous BS framework. 
By solving for the relativistic Salpeter wave functions, we evaluated the hadronic transition matrix elements 
as overlap integrals of the initial- and final-state wave functions,  providing a description of these three-body processes that incorporates nonperturbative contributions.

For the charmonium sector, our predictions for $\psi(2S) \to \chi_{cJ} e^+e^-$ agree well with the BESIII measurements. 
We have provided the first relativistic predictions for the $\psi(1D)$ (i.e., $\psi(3770)$) Dalitz decays. 
Unlike the $\psi(2S)$ case, all three muonic channels $\psi(1D) \to \chi_{cJ} \mu^+\mu^-$ ($J=0,1,2$) are kinematically allowed, although they are strongly phase-space suppressed. The branching fractions for the electron channels of $\psi(1D)$ are predicted to be at the $10^{-5}$ level, which is within the reach of upcoming BESIII analyses.

In the bottomonium sector, we studied the established $\Upsilon(2S)$ and the so-far unobserved $\Upsilon(1D)$ states. The branching fractions for $\Upsilon(2S) \to \chi_{bJ} e^+e^-$ are estimated to be of the order of $10^{-4}$, making them accessible at Belle~II and future super-$B$ factories. 
For $\Upsilon(1D)$, the predicted decay widths offer valuable theoretical input for the experimental search for this missing state.

A key feature of our approach is the consistent treatment of electromagnetic gauge invariance. 
For electron channels, gauge invariance was imposed to regulate the low-$Q$ region and ensure numerical stability, while for muon channels, the large lepton mass naturally regularizes the amplitude. 
This distinction highlights the importance of a rigorous treatment of the electromagnetic current in Dalitz decay studies.

In summary, our results establish a coherent theoretical basis for heavy quarkonium Dalitz decays, 
covering both charm and bottom flavors. Given that several predicted channels have branching fractions 
at the $10^{-4}$--$10^{-5}$ level, they fall within the sensitivity of current and upcoming experiments. 
We hope that these predictions will stimulate further experimental efforts at BESIII, Belle~II, and future tau--charm facilities.

\vspace{0.7cm} {\bf Acknowledgments}

This work was supported by the National Natural Science Foundation of China (NSFC) under the Grants No. 12575097 and No. 12375085. T. Wang was also supported by the Fundamental Research Funds for the Central Universities (2023FRFK06009). Q. Li was supported by the National Key R\&D Program of China\,(2022YFA1604803) and the Natural Science Basic Research Program of Shaanxi\,(No.\,2025JC-YBMS-020).

%\begin{thebibliography}
\bibliographystyle{modified-apsrev4-2}
\bibliography{ref}
%\printbibliography
%\end{thebibliography}

\end{document}